\documentclass[10pt,a4paper,twoside]{article}
\usepackage[a4paper]{geometry}
\usepackage[T1]{fontenc}
\usepackage{mathpazo}
\usepackage[latin1]{inputenc}							
\usepackage{amsmath,multicol}									
\usepackage{xcolor,graphicx}
\definecolor{bl}{rgb}{0.0,0.2,0.6}
\definecolor{nicered}{rgb}{.647,.129,.149}

\usepackage{hyperref}
\hypersetup{
    bookmarks=true,         
    pdftoolbar=true,        
    pdfmenubar=true,        
    pdffitwindow=false,     
    pdfstartview={FitH},    
    pdftitle={A statistical physics perspective on criticality in financial markets},    
    pdfauthor={Thomas Bury},     
    pdfnewwindow=true,      
    colorlinks=true,       
    linkcolor=bl,          
    citecolor=blue ,        
    filecolor=red!90,      
    urlcolor=blue ,          
}

\usepackage{sectsty}
\usepackage[compact]{titlesec}
\titleformat{\section}{\color{nicered}\large\bf}{\thesection}{1em}{}
\titleformat{\subsection}{\color{nicered}\normalsize\bf}{\thesubsection}{1em}{}
\titleformat{\subsubsection}{\color{nicered}\normalsize\bf}{\thesubsubsection}{1em}{}

\makeatletter							
\def\printtitle{
    {\color{bl} \flushleft \huge \@title\par}}		
\makeatother							

\title{Statistical pairwise interaction model of stock market}

\makeatletter							
\def\printauthor{
    {\hfill\parbox[b]{0.90\textwidth}{\flushleft \small \@author}}}				
\makeatother							

\author{%
	\textbf{\large Thomas Bury} \\[1\baselineskip]
    Service OPERA (CP194/5), Universit\'e libre de Bruxelles,\\
    Avenue F.D. Roosevelt 50, 1050 Brussels, Belgium\\
	Email:tbury@ulb.ac.be \\
	}

\usepackage{fancyhdr}
\usepackage{lastpage}	
\usepackage[runin]{abstract}		        
\abslabeldelim{:\quad}						%
\begin{document}
\printtitle

\printauthor

\begin{abstract}Financial markets are a classical example of complex systems as they comprise many interacting stocks. As such, we can obtain a surprisingly good description of their structure by making the rough simplification of binary daily returns. Spin glass models have been applied and gave some valuable results but at the price of restrictive assumptions on the market dynamics or others are agent-based models with rules designed in order to recover some empirical behaviours. Here we show that the pairwise model is actually a statistically consistent model with observed first and second moments of the stocks orientation without making such restrictive assumptions. This is done with an approach based only on empirical data of price returns. Our data analysis of six major indices suggests that the actual interaction structure may be thought as an Ising model on a complex network with interaction strengths scaling as the inverse of the system size. This has potentially important implications since many properties of such a model are already known and some techniques of the spin glass theory can be straightforwardly applied. Typical behaviours, as multiple equilibria or metastable states, different characteristic time scales, spatial patterns, order-disorder, could find an explanation in this picture.
\end{abstract}

\hrule
\footnotesize
\tableofcontents
\vspace{1em}
\hrule
\vspace{1em}
\normalsize

\section{Introduction}
\label{intro}

A highly interesting feature of complex systems is that sometimes the microscopic details of interactions are not necessary to explain the observed macroscopic structures (at least qualitatively). The most famous examples are the Ising and spin glasses models where the interactions are taken as constant or randomly distributed in a given neighbourhood.
It is amazing that the pairwise maximum entropy model also describes neural populations \cite{ref13}. This suggests that the most relevant properties ruling macroscopic behaviours of such complex systems are not the nature of microscopic entities but are order of interactions, their range and the topology.

One also finds collective phenomena in finance \cite{Dal,Jr}, non-random correlations \cite{refLaloux} and complex structures \cite{Mant,Onnela}.
Such phenomena can occur in systems compound by many interacting entities (where \emph{interaction} is taken at the larger sense of mutual influence).
Moreover, as recently observed \cite{Petra}, financial and neural networks have topological similarities (modular, hierarchical, small-world organization highlighted by an asset tree based approach). In this view, the spin-glass paradigm seems to be a seducing candidate to explain the market structure. Spin glasses were already applied to finance but with the restricting assumption that the market dynamics follows the soft-spins Langevin dynamics \cite{Rosenow}. There are also Ising like models which are agent-based models with specific rules such as \emph{"do what your neighbours do"} or more complex dynamical rules \cite{born,zhou,Sherr}. The latter approach is thus a different one that Rosenow's (or the present) approach where elementary entities are stocks and not traders (the most accessible observables are price returns).

The aim of this work is to show that market behaviour can be explained with no such hypothetical rules and that the aforementioned collective phenomena result from mutual influences of underlying constitutive entities, the stocks (in the same spirit of the characterization of collective phenomena in neural networks by neurons interaction without using other material than their activity time series \cite{ref13}).
We emphasize that this approach is a data-based approach. We do not introduce any rules or dynamical restriction. We only require the model fits first and second empirical moments. The reason is that the underlying microscopic details seem to be unnecessary to the macroscopic description of such phenomena. Indeed macroscopic behaviours in magnetic materials and in neural networks are consistently described by maximum entropy models even though electrons and neurons are undoubtedly completely different elementary entities at the individual scale (as well as their microscopic dynamics).
Furthermore, agent-based models can reveal interesting behavioural patterns but since such different dynamics as the neurons potential activity dynamics and spin dynamics can lead to the same macroscopic patterns, it seems natural to propose a complementary statistical and data-based approach allowing to relax almost any assumption.

Here, we consider stocks as economic entities influencing each other. The interaction process itself is not detailed. Instead, we propose a derivation of the pairwise model only based on the (incomplete) information embedded in the data without restricting assumption on an underlying dynamics. The only (rough) assumption that we made is prices binarizing to interpret daily movement as a bullish (or bearish) orientation. Such a simplification has already shown its power in neural networks and magnetic materials (at least in structure studies) where the complex interaction process is approximated by a pairwise model and the relevant variables (action potential and spin) are binarized.
In this work we provide evidence that an Ising model on a complex network can accurately describe the stock market. We show that almost all interaction strengths are Gaussian random variables, that Gaussian influences are compatible with non-Gaussian eigenvalues of the returns correlation matrix and that the mean influence scales as a power close to $-1$ of the system size. Furthermore frustration seems to be a key property since approximately half of the interaction strengths are negative.
We also propose an economic interpretation based on the mutual influence scheme developed in \cite{Brock}. Furthermore the interaction strengths can be thought as incentive since they are related to the Hessian matrix of the utility function \cite{mas}.

With these features, we conclude that the proposed model may fall into the class of spin glass exact mean field models. We also show that we can recover the largest (non-Gaussian) eigenvalue of the returns correlation matrix corresponding to the market eigenmode \cite{refLaloux}, making the link with the random matrix approach. This mapping and the first clue of the recovering of the market eigenmode suggest that a link to critical phenomena can be done in this paradigm.
Moreover the topological similarities between market and neural networks can find their origin in this common statistical model.
Other properties as the existence of hierarchical structures \cite{Mant,Onnela}, possibility of the order-disorder transition and synchronization \cite{Jr,Dal} can potentially be explained by the pairwise paradigm.

\section{The model}\label{sec:model}
\subsection{Inferred distribution}
\begin{figure}[!ht]
\begin{center}
\resizebox{\textwidth}{!}{%
  \includegraphics{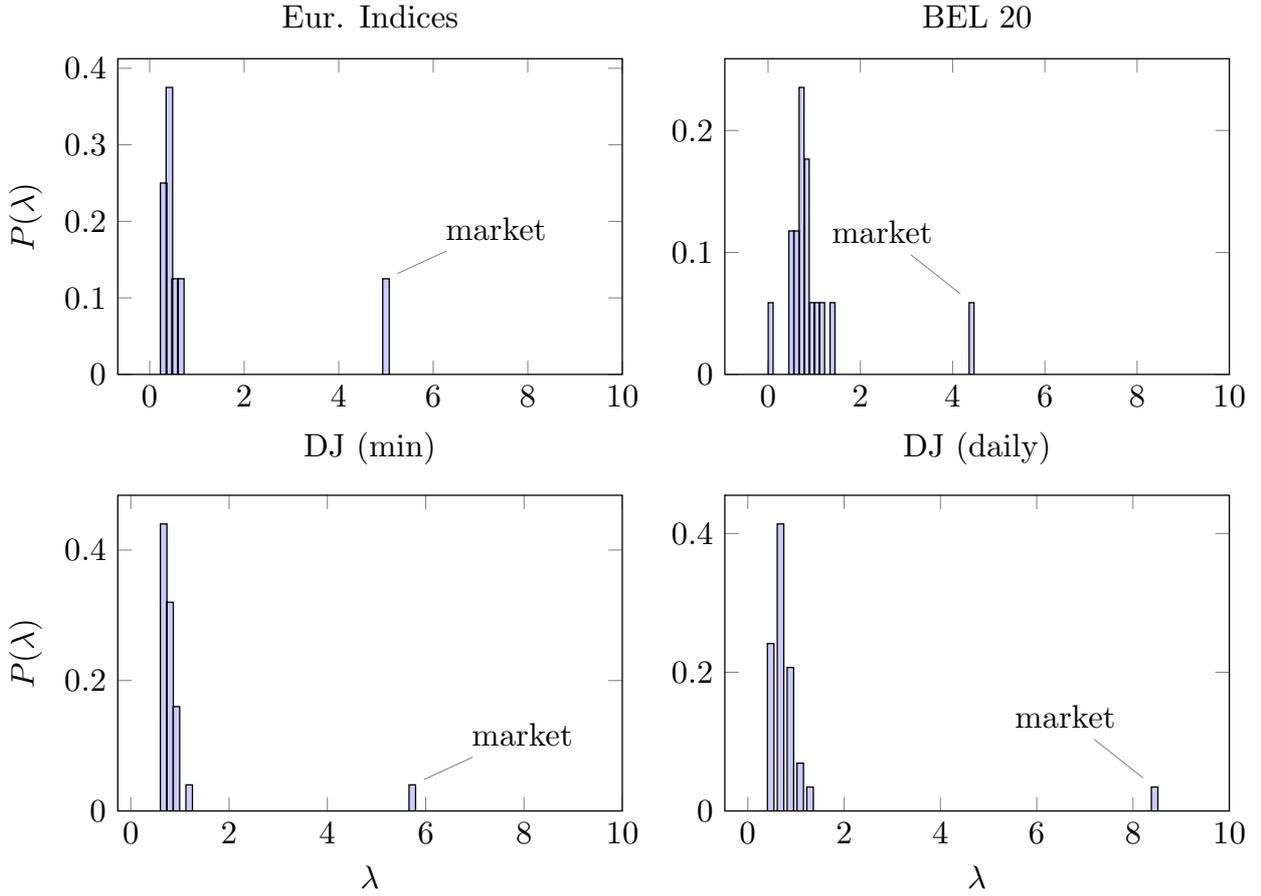}
}
\end{center}
\caption[Indices eigenmode]{Probability distribution of eigenvalues of the binarized returns correlation matrix for 8 European indices (top-left), Bel20 (top-right), Dow Jones at minute sampling (bottom-left) and Dow Jones daily (bottom-right). The market-mode is pinned.}
\label{fig:DJmode}
\end{figure}

Our aim is to set up a model describing the market state and its structure based only on statistical considerations. This requires a way to infer the probability distribution in order to get the observables (here, the associated moments). The model will also allow the study of the market structure. All these quantities will be defined below.
We consider a set of $N$ market indices or $N$ stocks with binary states $s_{i}$ ($s_{i}=\pm1$ for all $i=1,\cdots,N$). A system configuration will be described by a vector $\textbf{s}=(s_{1},\cdots,s_{N})$. The binary variables will be equal to $1$ if the associated closing price is larger than (or equal to) the opening one and equal to $-1$ if not. We choose open-to-close rather than close-to-close returns to avoid over-night effect and the weekend gap (Friday-Monday closings). A configuration $\textbf{s}$ is a binary version of stock returns. Such a simplification of returns is made to study the market structure and will be justified a posteriori if the results are consistent with the data. A first clue that is not a too rough approximation is that it preserves the market eigenmode (largest eigen-value of the price-returns covariance matrix) \cite{refLaloux} as illustrated in Fig-\ref{fig:DJmode}

Another motivation of this approximation is that the resulting binary pairwise model allows collective phenomena which are observed in the market. We will discuss the description of the collective phenomena (structure reorganization, synchronization, etc.) by this pairwise model in a forthcoming work.

We seek to establish the less structured model explaining only the measured mean orientations $q_{i}$ and instantaneous pairwise correlations $q_{kl}$ in terms of theoretical moments $\langle s_{i}\rangle$ and $\langle s_{k}s_{l}\rangle$ without making any further assumption. The brackets $\langle\cdot\rangle$ denote the average with respect to the unknown distribution $p(\textbf{s})$. As the entropy of a distribution measures the randomness or lack of interaction among binary variables, a way to infer such probability distribution knowing the mean orientations and correlations is the maximum entropy principle (MEP). Jaynes showed how to derive the probability distribution using the maximum entropy principle \cite{ref12}. It consists in the following constrained maximization

\begin{eqnarray}\label{maxent}
  & &  \max_{\substack{\{p(\mathbf{s})\}}} S(\textbf{s})= \max_{\substack{\{p(\mathbf{s})\}}}\left\{ -\sum_{\{\textbf{s}\}}p(\textbf{s}) \,\ln p(\textbf{s})\right\}  \\ \nonumber
   &\mathrm{s.t}&  \sum_{\{\textbf{s}\}}p(\textbf{s})=1,\quad \sum_{\{\textbf{s}\}}p(\textbf{s})s_{i}=q_{i},
   \quad \sum_{\{\textbf{s}\}}p(\textbf{s})s_{i}s_{j}=q_{ij} \nonumber
\end{eqnarray}

The resulting  2-agents distribution $p_{2}(\textbf{s})$ is the following

\begin{equation}
p_{2}(\textbf{s})=\mathcal{Z}^{-1}\exp\left(\frac{1}{2}\sum_{i, j}^{N}J_{ij}s_{i}s_{j}+\sum_{i=1}^{N}h_{i}s_{i}\right)\equiv\frac {e^{- \mathcal{H}(\textbf{s})}}{\mathcal{Z}}\label{Lagrange}
\end{equation}

where $J_{ij}$ and $h_{i}$ are the Lagrange multipliers and $\mathcal{Z}$ a normalizing constant (the partition function). They can be expressed in terms of partial derivatives of the entropy as

\begin{equation}
\frac{\partial S(\textbf{s})}{\partial q_{i}} = -h_{i} \qquad
\frac{\partial S(\textbf{s})}{\partial q_{ij}} = -J_{ij}\label{multipliers}
\end{equation}

Thus preferences are conjugated to mean orientations and pairwise influences to pairwise correlations.

Cumulants are obtained from this model and we give their relation to the interaction strengths. As the statistical model (\ref{Lagrange}) is expressed as a Gibbs distribution, we have the relations

\begin{equation}\label{moments}
\langle s_{i_{1}}\ldots s_{i_{N}}\rangle_{\mathrm{c}}=\partial^{N}\ln \mathcal{Z}/\partial h_{i_{1}}\ldots \partial h_{i_{N}}
\end{equation}

where $\langle\cdot\rangle_{\mathrm{c}}$ is the cumulant average \cite{Kubo}. This relation gives the link between $\mathbf{J}$ and pairwise correlations. If the partition function $\mathcal{Z}$ cannot be explicitly computed, we can use Plefka series \cite{Plef} or a variational cumulant expansion \cite{Barb}.

\begin{figure}[!ht]
\begin{center}
\resizebox{0.75\textwidth}{!}{%
  \includegraphics{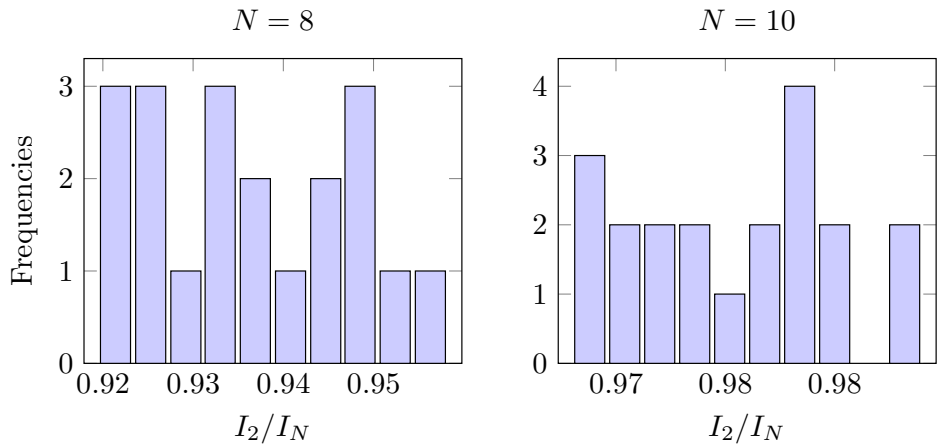}
}
\end{center}
\caption{Multi-information ratio $I_{2}/I_{N}$ for 20 sets of 8 randomly chosen stocks (left) and for 20 sets of 10 randomly chosen stocks. Sampling time-scale is the minute, the sample length is $3\times 10^{4}$ points and parameters were estimated with a regularized pseudo-maximum likelihood method.}
\label{fig:MI}
\end{figure}

Finally, we test if higher order influences should be ruled out. We proceed by using the multi-information criterion \cite{ref16,ref13}. We sketch here the basic idea of this criterion. Considering a financial network of $N$ entities, one can obtain maximum entropy distributions $p_{k}(\textbf{s})$ which are consistent with $k$th-order correlations (for any $k=1,\cdots,N$) like in (\ref{maxent}). The case $k=N$ is an exact description of the financial network. Thus the entropies $S_{k}=S[p_{k}]$ of these distributions decrease with increasing $k$ toward the true entropy $S=S[p_{N}]$ since more correlation reduces the entropy. The multi-information $I_{N}\equiv D_{\mathrm{KL}}\left(p_{N}||p_{1}\right)$ is a measure of the total amount of correlations in the system. Thus if the ratio $I_{2}/I_{N}=(S_{1}-S_{2})/(S_{1}-S_{N})$ is close to $1$ then pairwise correlations provide an effective description of the correlation structure (Where $ D_{\mathrm{KL}}$ is the Kullback-Leibler divergence).
For a set of 8 European indices, we obtain $I_{2}/I_{N}=98.2\%$ which means that pairwise correlations represent most of correlations. For the Dow Jones (minute sampling time-scale and $3\times10^{4}$ points), we obtain $I_{2}/I_{N}=95.7\%$ in average. In the latter case we consider 20 sets of 8 randomly chosen stocks and 20 sets of 10 randomly chosen stocks (values for which direct sampling of the distribution gives a good estimate); the results are illustrated in Fig-\ref{fig:MI}.

\subsection{Interpretation}

The Gibbs distribution (\ref{Lagrange}) is similar to those given by Brock and Durlauf in the discrete choice problem \cite{Brock} and in stochastic  models in macroeconomics \cite{Aoki}, but also to the Ising model used in description of magnetic materials and neural networks \cite{ref8,ref13}. This is also a special case of Markov random fields \cite{ref15}. We emphasize that the Gibbs distribution and the concept of information entropy naturally arise from stochastic modelling in economics. This is discussed at length in \cite{Aoki}.
We interpret the objective function $\mathcal{H}(\textbf{s})$ defined by the MEP as follows.
Pairwise interactions between economic agents are modelled by interaction strengths $J_{ij}$ (which describe how $i$ and $j$ influence each other). They can be thought as a measure of the degree of co-movement (coherence) of a time-series pair. As possible underlying causes of those interactions, we may thus think to the economic background, company management, traders strategies, etc. This should be investigated in an econometrical study.
The interaction matrix $\textbf{J}$ is set to be symmetric in this first approach. There is disagreement or conflict between entities when the weighted product of their orientations $J_{ij}s_{i}s_{j}$ is negative. If two shares are supposed to move together ($J_{ij}>0$), a conflicting situation is the one where they do not have the same orientation (bearish or bullish).

We include idiosyncratic preferences or individual biases of stocks, here the willingness to be bullish or not. These Lagrange multipliers $h_{i}$ can also be interpreted as external influences on entities $i$ induced by the macroeconomic background. By example a company can prosper and make benefits during a crisis period and the associated stock can still fall simultaneously because investors are negatively influenced by the economic background. The stock will have a propensity to fall even if profits are made. If the orientation of the stock satisfies its preference, $h_{i}s_{i}$ will be positive. The total conflict of the system is then given by

\begin{equation}\label{Hfunc}
\mathcal{H}(\textbf{s})=-\frac{1}{2}\sum_{i=1}^{N}\sum_{j=1}^{N}J_{ij}\, s_{i}s_{j}-\sum_{i=1}^{N}h_{i}s_{i}
\end{equation}

We interpret $\mathcal{H}(\textbf{s})$ as the opposite of the so-called utility function $\mathcal{U}(\textbf{s})=-\mathcal{H}(\textbf{s})$ with a pairwise interacting and idiosyncratic parts \cite{Brock}. Consequently interaction strengths can be viewed as incentive complementarities. Indeed we have $\partial^{2} \mathcal{U}/\partial s_{i}\partial s_{j}=J_{ij}$ . The larger $J_{ij}s_{i}s_{j}$, the stronger the strategic interaction between $i$ and $j$.

We emphasize that this Ising like model is forced upon us as the statistically consistent model with measured orientations and correlations. It is not an analogy based on specific hypotheses about the market dynamics and it necessarily implies a multivariate picture of the markets as it should be.

\subsection{Parameters estimation}

The parameters $\{J_{ij},h_{i}\}$ can potentially be exactly computed by performing explicitly the maximization (\ref{maxent}) so that the theoretical moments $\langle s_{i}\rangle$ and $\langle s_{i}s_{j}\rangle$ match the empirical ones $q_{i}$ and $q_{ij}$. This method requires the computation of $2^N$ terms. If this number is too large, the computation is unfeasible and we can benefit from one of the methods described in \cite{ref4}. The parameters should be valued such that the constraints are satisfied in (\ref{maxent}). Generally, redrawing the parameters from their distribution will lead to wrong values of the first and second moments. Therefore knowing only the functional form of the distribution is insufficient, we must know their exact values. In this paper we use a second order mean-field inversion \cite{ref4} (consistently with our results). Generally this inversion method requires ten or so entities and a sample size $T$ larger than the number of entities $N$. In the following we have $T>20N$ and $N>10$.
This inversion technique, to infer interaction strengths from data, is based on the following relation ($i\neq j$)

\begin{equation}\label{inversion}
    (\mathbf{C}^{-1})_{ij}=-J_{ij}-J_{ij}^{2}\,q_{i}q_{j}
\end{equation}

Given the relation (\ref{inversion}), if data are noise dressed, the inferred interaction matrix will also be noise dressed. Moreover, as the proposed model is a maximum entropy model, the parameters should be adjusted to satisfy the constraints in (\ref{maxent}). Thus any inversion method will be noise sensitive. Lastly, we note that the MEP is also sample-dependent since Lagrange multipliers are fitted to recover the first and second moments. It does not necessarily mean that $J_{ij}$ are time-dependent but it seems intuitive that they are actually time-dependent since a company can die out, be restructured or removed from its index.

\section{Mean field mapping}\label{sec:consistency}

The previous model can be thought as an Ising spin glass on a complex network \cite{Dor}. Indeed the objective function of this model (equivalent to the Ising Hamiltonian) can be rewritten as

\begin{equation}
\mathcal{H}(\textbf{s})=-\frac{1}{2}\sum_{i=1}^{N}\sum_{j=1}^{N}J_{ij}A_{ij}\, s_{i}s_{j}-\sum_{i=1}^{N}h_{i}s_{i}
\end{equation}

where $A_{ij}$ are the entries of the adjacency matrix, equal to one if the nodes $i$ and $j$ are connected and equal to zero if they are not. For a complete graph ($A_{ij}=1$ for all pairs) Thouless-Anderson-Palmer (TAP) equations are exact if the number of nodes tends to infinity and if the $J_{ij}$ are independent and identically distributed (IID) gaussian random variables with mean and variance scaling as $N^{-1}$ \cite{Dor,Plef}. We can check if observed mean orientations are well approximated by TAP equations

\begin{equation}\label{TAP}
\langle s_{i}\rangle_{\mathrm{c}}=\tanh\left(h_{i}+\sum_{j}J_{ij}\langle s_{j}\rangle_{\mathrm{c}}-\sum_{j}J_{ij}^{2}\langle s_{i}\rangle_{\mathrm{c}}[1-\langle s_{j}\rangle_{\mathrm{c}}^{2}]\right)
\end{equation}

\begin{figure}[!ht]
\begin{center}
\resizebox{.75\textwidth}{!}{%
  \includegraphics{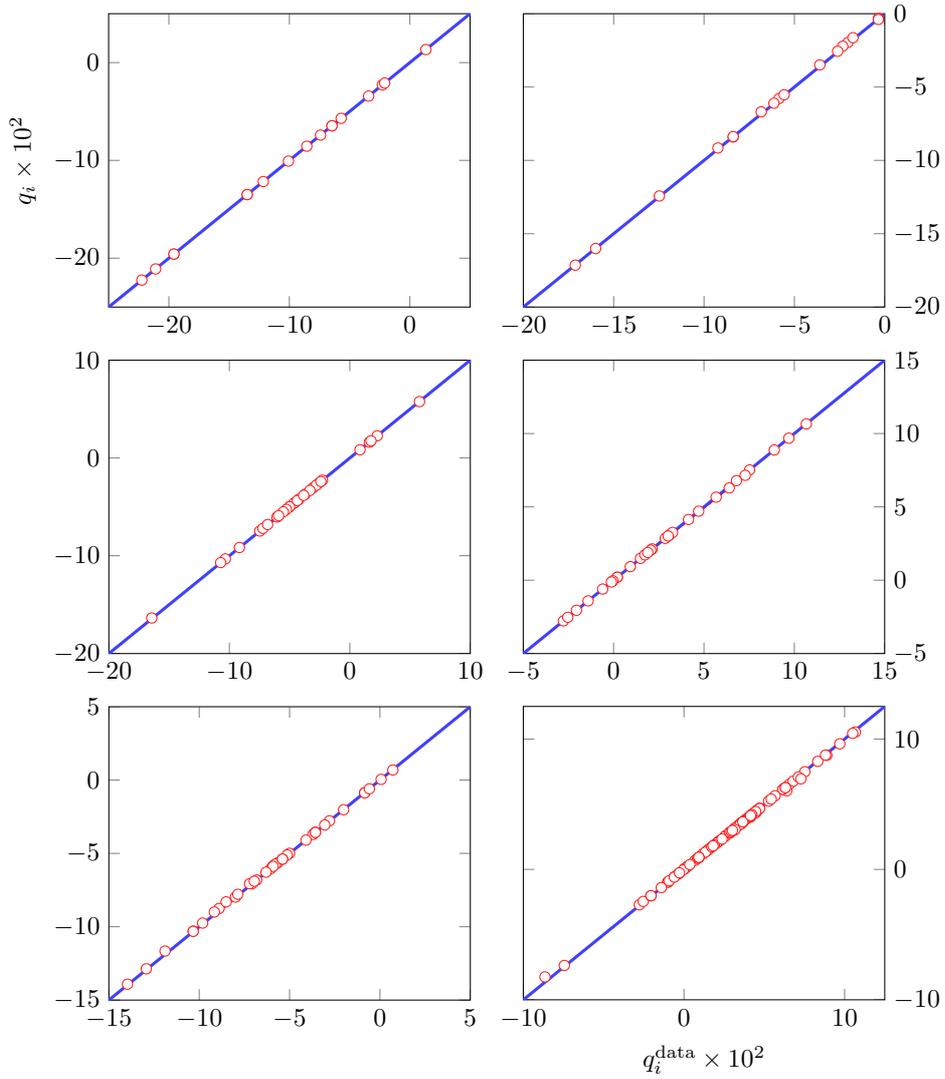}
}
\end{center}
\caption{Comparison of TAP mean orientations (circles) and empirical ones. The straight line shows equality. Respectively from top left to bottom right (with increasing system size): BEL20, AEX, DAX, DJ, CAC, S$\&$P100.}
\label{fig:choices}
\end{figure}

Below, we show that first and second empirical cumulants are indeed well approximated by TAP mean-field for different market indices and for different system sizes. We consider the $N$ stocks of the BEL20, AEX, DAX, Dow Jones, CAC40 and S$\&$P100 indices respectively during $T=1050$, $T=1400$, $T=1550$ $T=2500$, $T=1550$ and $T=2500$ trading days, such that $T\gg N$ (a trading year is usually about 250 trading days). All these data can be downloaded from the web site Yahoo! Finance \cite{ref3}. We compute TAP mean orientations of each stock in this large time window and we compare them with their empirical mean values. The results are illustrated in Fig-\ref{fig:choices}.

TAP mean orientations are indeed a good description of empirical mean orientations, the typical relative deviation is less than $1\%$. As a further test, we also compare empirical variances of orientations to their TAP values. Variances of orientations are $\langle s_{i}^{2}\rangle_{\mathrm{c}}=1-\langle s_{i}\rangle_{\mathrm{c}}^{2}$ inserting the TAP approximation leads to $\langle s_{i}^{2}\rangle_{\mathrm{c}}=1-\tanh^{2}(h_{i}+\sum_{j}J_{ij}\langle s_{j}\rangle_{\mathrm{c}}-\sum_{j}J_{ij}^{2}\langle s_{i}\rangle_{\mathrm{c}}[1-\langle s_{j}\rangle_{\mathrm{c}}^{2}])$.


Variances are also well approximated by TAP variances, the typical relative deviation is about $1\%$. Using error propagation, one can evaluate the error on the estimation of
third order cumulants $\langle s_{i}^{3}\rangle_{\mathrm{c}}=2(\langle s_{i}\rangle_{\mathrm{c}}^{3}-\langle s_{i}\rangle_{\mathrm{c}})$ and higher order cumulants which are expressed in terms of TAP orientations.
The TAP mean field method is exact, in the so-called thermodynamic limit $N\rightarrow\infty$, for the infinite-range interactions provided that the following condition is satisfied \cite{Plef}

\begin{equation}\label{TAP2}
x\equiv 1-(1-2Q_{2}+Q_{4})>0 \quad \mathrm{with} \quad Q_{\nu}=N^{-1}\sum_{i=1}^{N}q_{i}^{\nu}
\end{equation}

We checked that this condition is fulfilled for each of the previous data sets and so our use of TAP equation was justified. We showed that an Ising model on a complex graph can accurately describe the stock market for different and typical system sizes as TAP equations give results consistent with the data.

The good adequation between empirical and TAP cumulants suggests that the market network should be like a complete graph, with pairwise influences which should be Gaussian ones and scale as the inverse of the system size. However, the real financial network may be not actually a complete graph even if the only null entries of the interaction matrix are the diagonal ones. Indeed one knows that a part of the correlations is noise \cite{refLaloux}. Moreover, finite size sample also implies errors in the parameters estimation.
It would be nice if, in addition, the interaction matrix entries $J_{ij}$ were actually gaussian random variables as needed by the TAP mean-field approach. This would make the link with the Gaussian spin glass theory \cite{ref8}. We want to emphasize that one should not confuse the interaction matrix with the covariance matrix of the returns. The fact that $\textbf{J}$ entries are normally distributed does not mean that there are only noisy movements in the market. The $\textbf{J}$ matrix describes the pairwise interactions, not directly the correlations.

\begin{figure}[!ht]
\begin{center}
\resizebox{\textwidth}{!}{%
  \includegraphics{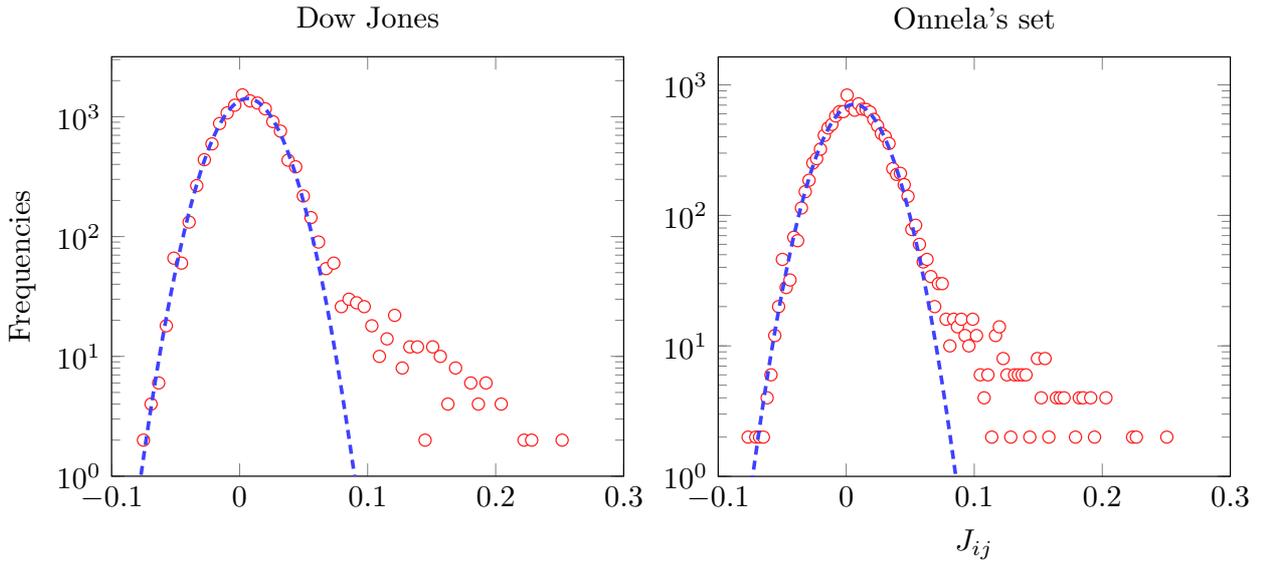}
}
\end{center}
\caption{Top: Empirical frequencies of pairwise influences for the DJ (minute time-scale) and bottom: the Onnela's set . The dashed line is a Gaussian fit of the influences frequencies distribution amputated of its upper tails.}
\label{fig:distr}
\end{figure}

\begin{figure}[!ht]
\begin{center}
\resizebox{\textwidth}{!}{%
  \includegraphics{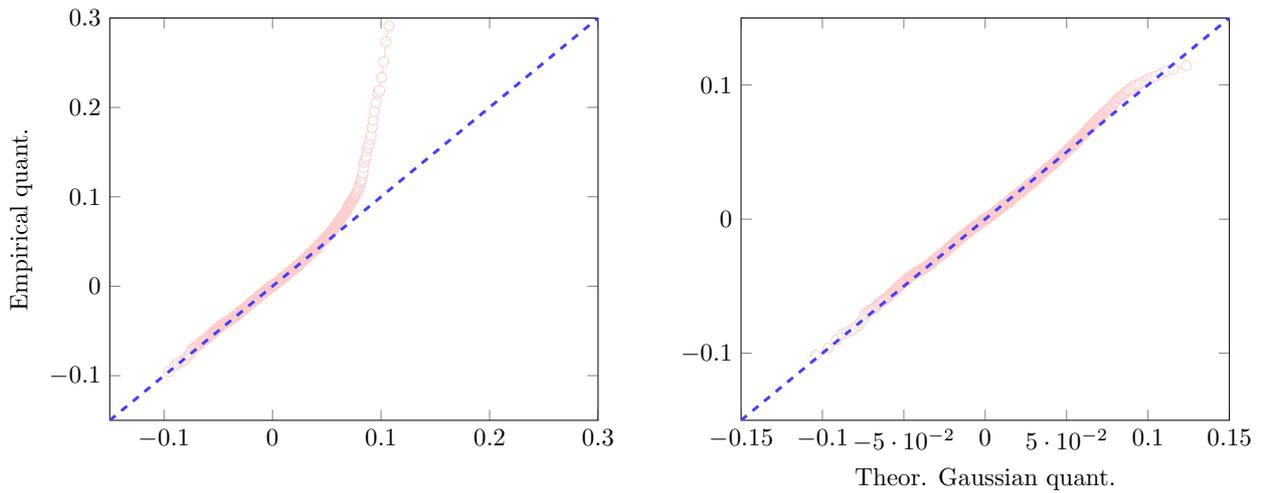}
}
\end{center}
\caption{Comparison of S$\&$P100 empirical 1000-quantiles (circles) and theoretical ones. The straight line shows equality. Respectively from left to right: all the $4950$ entries of the $\mathbf{J}$ matrix and the results without the last 200 entries.}
\label{fig:quant}
\end{figure}

We illustrated in Fig-\ref{fig:distr} the empirical frequencies of the estimated mutual influences. We consider the CAC index and a set of 116 NYSE stocks observed during 4800 trading days (available at \url{www.jponnela.com}). The frequencies distribution does not seem to be exactly Gaussian since the upper tail is fatter than in the Gaussian distribution. To formalize this observation, we first use a qualitative normality test. We compare the empirical 1000-quantiles (permilles) with the theoretical gaussian 1000-quantiles. If the $J_{ij}$ are Gaussian random variables, we should obtain a linear relation between these both quantities. We illustrated our results in Fig-\ref{fig:quant}.

We tested the normality of the interaction strengths for the previous six market indices. We obtained similar results than those illustrated in Fig-\ref{fig:quant}. The upper tail of the empirical distribution is also found fatter than the Gaussian one but the bulk of the distribution seems to be Gaussian.
Then we use the $\chi^{2}$ and the Jarque-Bera statistical normality tests on the $\mathbf{J}$ upper triangular part amputated of its upper tail. They do not lead to the rejection of the null hypothesis that the bulk of the underlying distribution is a Gaussian one.

Last, to evaluate the importance of the noise in the estimation, we simulate the binary time-series (for different sizes and sample lengths) with the maximum entropy conditional flipping probability $p(s_{i,t}=-s_{i,t-1}|\mathcal{H}_{t})$ given the state at time $t$. The influence matrix was taken homogenous with all entries equal to the empirical mean $\bar{J_{ij}}$ of the considered index in those simulations. We then estimate the influence matrix with those artificial data. Ideally, the standard deviation of estimated artificial influences $\sigma_{\texttt{noise}}$ should be much smaller than the one of real influences $\sigma_{J}$. The results are reported in Table-\ref{tab:noise}. Depending on the sample length, the noise seems to be significant but not the dominant part of the estimation excepted for large system size.

\begin{table}[!ht]
\caption{Quantification of noisy part of the variance of inferred mutual influences.}
\label{tab:noise}
\begin{center}
\begin{tabular}{lcr}
\hline
Index         & sample length ($T$) & $\sigma_{\texttt{noise}}/\sigma_{J}$\\ \hline
AEX(daily)    &  $1.4\times 10^{3}$ & 0.22                                \\
DJ(min)       &  $3.0\times 10^{4}$ & 0.24                                \\
DJ(daily)     &  $2.5\times 10^{3}$ & 0.31                                \\
Onnela(daily) &  $4.8\times 10^{3}$ & 0.74                                \\
Cac(daily)    &  $1.5\times 10^{3}$ & 0.75                                \\
\hline
\end{tabular}
\end{center}
\end{table}

\begin{figure}[!ht]
\begin{center}
\resizebox{0.75\textwidth}{!}{%
  \includegraphics{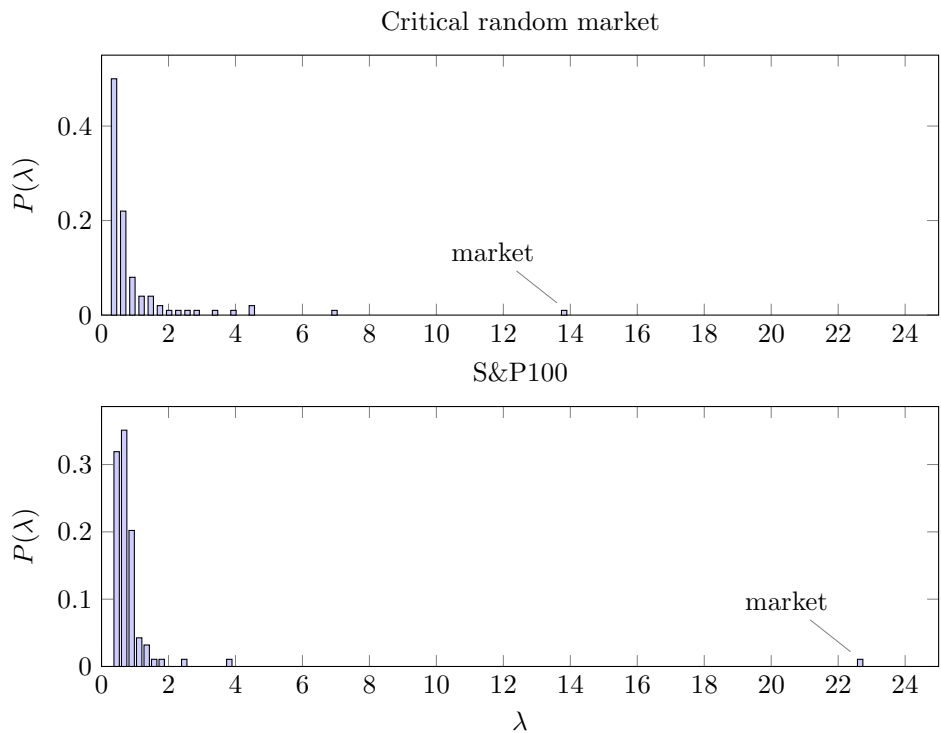}
}
\end{center}
\caption{(Top) Typical probability distribution of returns covariance matrix eigenvalues at the transition. A \emph{critical} random market is able to exhibit non-Gaussian covariance matrix. (Bottom) The empirical probability distribution of the covariance matrix eigenvalues of the S$\&$P100 index.}
\label{fig:crit}
\end{figure}

However it is not obvious if the upper tail can be neglected or not (one knows that one cannot neglect the non-Gaussian part of the correlation matrix). The non-Gaussian part of the distribution may also be an inference artefact (since less than $10\%$ of the influences are non-Gaussian ones).
We are tempted to let the door open to the case of Gaussian influences. Indeed, in addition to the previous evidence of TAP matching, Gaussian interactions are compatible with the observed market eigenmode. Consider the simplest situation where $J_{ij}$ are really \emph{IID} Gaussian random variables with zero mean (thus including the frustration since half of the pairwise influences are negative). The largest eigenvalues of the returns covariance matrix are linked to eigenvalues of the $\textbf{J}$ matrix by the relation $[1-J_{\lambda}+J^2]^{-1}$ in the mean field approach, where $J_{\lambda}$ is an eigenvalue of the $\mathbf{J}$ matrix and $J^{2}\equiv N\,\mathrm{VAR}(J_{ij})$ \cite{ref8}. This quantity is large when $J_{\lambda}$ lies in the vicinity of $1+J^{2}$. In this particular situation the largest eigenvalue of the interaction matrix is equal to $2J$. A special case is the one where $J=1$ which corresponds to the transition in the Sherrington-Kirkpatrick model. The largest eigenvalue of the covariance matrix diverges in the limit of infinite number of entities. We illustrated this behaviour for $N=100$ interacting stocks with IID Gaussian interaction strengths in Fig-\ref{fig:crit}.

In the present applications entries of the interaction matrix do not seem to have a common mean and variance; therefore the relation between both kinds of eigenvalues is more complex than the former one. It is then non-obvious to conclude wether the interaction strengths are actually Gaussian or whether the right fat-tail of their distribution is an actual deviation to the normal distribution (and not an inference artifact). The possible interpretation of a market behaving as a critical complex system should be investigated in detail.

The possible normality of interactions has another consequence: the $\mathcal{U}(\mathbf{s})$ function defines a Gaussian process. Our model is thus a random utility model and tools of the random matrix theory \cite{refCar} can be useful to study the market structure, as they already are in the study of stock return correlations \cite{refLaloux}. We also checked that a significant part of interaction strengths are negative ($37.2\%$ for the S$\&$P100 and $24.3\%$ for the Dow Jones). Together with the former observation of a possible market mode even with truly Gaussian $J_{ij}$, we may think that the  frustration is a main feature of the market interaction structure. We may think the frustration as competitive influences between cyclic sectors (more correlated to the global health of the worldwide economy and thus privileged by the investor during a growth period) and the defensive sectors.


%

\begin{figure}[!ht]
\begin{center}
\resizebox{0.6\textwidth}{!}{%
  \includegraphics{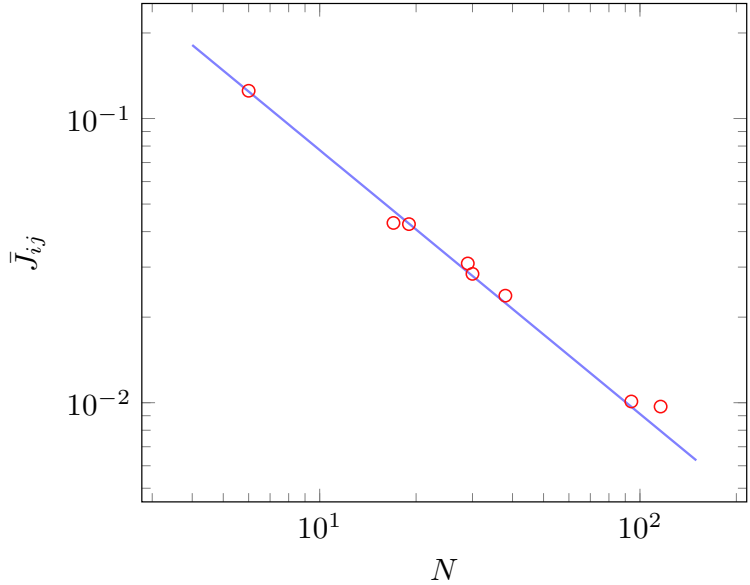}
}
\end{center}
\caption{Log-log plot of the mean interaction strengths in function of the typical system sizes (circles). The straight line is a non-linear fit (power-law).}
\label{fig:MeanJ}
\end{figure}

Another main feature is the scaling of the mean interaction strengths as a function of the system size, as needed in the TAP approach. To ensure that the $\mathcal{H}$ function (\ref{Hfunc}) is extensive (scaled as $\mathcal{H}\propto N$), the mean strength $\bar{J_{ij}}$ should be scaled as $\bar{J_{ij}}\propto N^{-1}$ \cite{binder}. Hereafter, we show that mean interaction strengths exhibit indeed these scaling properties for characteristic system sizes encountered in stock markets. We infer interaction strengths on a common time window of 1000 trading days (four years long time series) for the following indices (given in increasing size): BEL20, AEX, DAX, DJ, CAC40, S$\&$P100 and Onnela's set.
We add a supplementary point by computing the interaction strengths between six major European indices (adding another order of magnitude of the typical system size). The results are illustrated in Fig-\ref{fig:MeanJ}

We adjust a power law $aN^{-\alpha}$ to the data (illustrated by a straight line in a log-log graphic). The resulting coefficient of determination $R^{2}=0.997$. The estimation of the slope is $\hat{\alpha}=0.928\pm0.030$ (mean $\pm$ s.d). We conclude from this analysis that indeed the mean strength scale as $\bar{J_{ij}}\propto N^{-\alpha}$ with alpha close to $1$, in the interval of characteristic system sizes encountered in financial markets. This implies that the utility function (\ref{Hfunc}) may be an extensive one and thus that the quantities which derive from this function may be correctly scaled.
We note this is not the case for neural networks where the typical interaction strengths seem to be constant for growing $N$. In a physical system this situation is equivalent to lowering the temperature (leading to a \emph{frozen} state). The scaling $\bar{J_{ij}}\propto N^{-1}$ implies on the contrary that financial systems will not freeze and will not have the error-correcting property \cite{ref13}.

\begin{figure}[!ht]
\begin{center}
\resizebox{0.75\textwidth}{!}{%
  \includegraphics{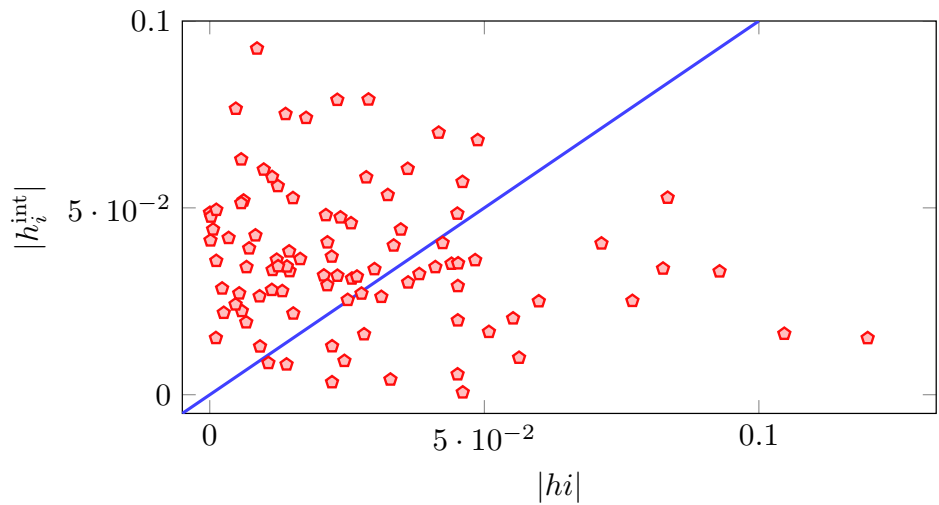}
}
\end{center}
\caption{Comparison of the S$\&$P100 index internal bias $h_{i}^{\mathrm{int}}$ experienced by a stock versus its individual bias $h_{i}$. In the upper left triangle, the internal bias dominates the intrinsic bias. Similar results are obtained even for smaller indices (like the BEL20 or AEX).}
\label{fig:bias}
\end{figure}

Since interaction strengths can be weak, we may ask if they have actually a predominant role in the market structure or if the values of interesting quantities are principally determined by individual bias $h_{i}$. From the relation (\ref{Lagrange}) we conclude that the orientation of each stock $s_{i}$ is subjected to a total bias $h_{i}+2^{-1}\sum_{j}J_{ij}s_{j}$. Interactions play a key role if the internal bias $h_{i}^{\mathrm{int}}=2^{-1}\sum_{j}J_{ij}s_{j}$ is significant compared to the individual bias $h_{i}$. We checked that they are in average of the same magnitude order. The results for the S$\&$P100 index are illustrated in Fig-\ref{fig:bias}.

Collecting previous results, we gave some empirical evidences that the financial market is described by a statistical model equivalent to an infinite range mean field spin glass. The spin glass theory provides an effective toolbox to study the financial markets structure as a complex system \cite{ref8,binder}.

However, we do not identify this statistical model to the Sherrington-Kirkpatrick model because there is no guarantee that interactions are quenched (static mean and variance) or even drawn from the same distribution. If the parameters are not quenched, their values can possibly change before the equilibration (if there is any) of the system.

\section{Conclusion}

We provided empirical evidences that the financial network is accurately described by a statistical model which can be thought as an Ising model on a complex (possibly complete) graph with scaled interaction strengths. This results lays down the pairwise model as a consistent paradigm in the study of stock market since first and second order influences are the dominant ones.
In particular, we showed that orientations are accurately inferred by the TAP equation (in the stability domain). Linked to this result, we checked that almost all the interaction strengths are Gaussian random variables, their average values scale as $N^{-\alpha}$ with $\alpha$ close to $1$. A significant part of the interaction strengths are negative, leading to frustration. Moreover, we showed that this model with truly Gaussian and scaled ($N^{-1}$) influences is able to recover the market eigenmode.
Consequently the proposed model may be thought as an exact mean-field one and the market state cannot be deduced by an observation of a small part of it.
Some methods developed in the spin glasses and neural networks theories could be applied in the study of the financial network, but we must pay attention to the specificities of each discipline, like the characteristic system size and the scaling of interactions for instance.
Some of the consequences are the existence of metastable states, the emergence of collective phenomena and spatial patterns, etc.
Furthermore, the processes taking place in the stock market should then occur at different timescales. The finite size of the stock market avoids the thermodynamic limit even as an approximation. Indeed the characteristic index size is about $N=10^2$ or $N=10^3$, much smaller than in physical or biological systems. Even if the relevant variables are correctly scaled, the fluctuations can be significant because at equilibrium they typically scale as $\sqrt{N}$ (far from transition).

Other potentialities could be the clustering analysis, the characterization of the financial network (confirming the small-worldness and scale-freeness within this framework), the study of crises through the interaction matrix and Monte-Carlo simulations. Some of these aspects will be investigated in further works.

\section*{acknowledgement}
I would like thank B. De Rock, P. Emplit and A. Smerieri for their helpful comments and discussions. This work was undertaken with financial support from the Solvay Brussels School of Economics and Management.
%

\end{document}